\documentclass[10pt]{article}

\usepackage{amsmath}
\usepackage{amssymb}

\usepackage{graphicx}

\usepackage{cite}

\usepackage{color} 

\newcommand{\figref}[1]{Figure~\ref{fig:#1}}
\newcommand{\figlabel}[1]{\label{fig:#1}}

\newcommand{\tabnum}[1]{\ref{tab:#1}}
\newcommand{\tabref}[1]{Table~\tabnum{#1}}
\newcommand{\tablabel}[1]{\label{tab:#1}}

\usepackage[labelfont=bf,labelsep=period,justification=raggedright]{caption}

\makeatletter
\renewcommand{\@biblabel}[1]{\quad#1.}
\makeatother

\date{}

\begin{document}

\begin{flushleft}
{\Large
\textbf{CONCOCT: Clustering cONtigs on COverage and ComposiTion}
}
\\
Johannes Alneberg$^{1,*}$, 
Brynjar Sm\a'ari Bjarnason$^{1,*}$, 
Ino de Bruijn$^{1,2}$, 
Melanie Schirmer$^3$, 
Joshua Quick$^4$, 
Umer Z. Ijaz$^3$, 
Nicholas J. Loman$^4$, 
Anders F. Andersson$^{1,\dagger}$ \& Christopher Quince$^{3,\dagger}$
\\
 \bf{1} KTH Royal Institute of Technology, Science for Life Laboratory, School of Biotechnology, Division of Gene Technology, Stockholm, Sweden
\\
 \bf{2} BILS, Bioinformatics Infrastructure for Life Sciences,
Vetenskapsr\r{a}det, Stockholm, Sweden
\\
 \bf{3} University of Glasgow, Glasgow, UK
\\
 \bf{4} Institute of Microbiology and Infection, University of Birmingham, Birmingham, UK
\\
\noindent $^{*}$ Joint first authors \\
\noindent $^{\dagger}$ Correspondence and requests for materials
should be addressed to A.F.A~(email: anders.andersson@scilifelab.se) or C.Q.~(email: christopher.quince@glasgow.ac.uk)
\end{flushleft}

\section*{Abstract}
Metagenomics enables the reconstruction of microbial genomes in complex microbial communities without the need for culturing. Since assembly typically results in fragmented genomes the grouping of genome fragments (contigs) belonging to the same genome, a process referred to as binning, remains a major informatics challenge. Here we present CONCOCT, a computer program that combines three types of information - sequence composition, coverage across multiple sample, and read-pair linkage - to automatically bin contigs into genomes. We demonstrate high recall and precision rates of the program on artificial as well as real human gut metagenome datasets.
\section*{Introduction}

Metagenomic shotgun sequencing of microbial communities is widely used to investigate the genetic potential and taxonomic composition of microbial communities. It has also been used to reconstruct the genomes of individual microbial populations\cite{tyson04,herlemann13}, and with the rise of high-throughput sequencing this has become possible for highly complex environments such as sediments and soils\cite{kantor13}. Repeated sequences within and across genomes, limited coverage, sequencing errors, and strain-level variation in gene content all result in fragmented genomes following genome assembly. A major bioinformatics challenge in metagenomics is therefore the process of binning contigs into species-level groups. If closely related reference genomes are lacking, binning has to be conducted in an unsupervised fashion. While numerous binning methods exist, few combine multiple lines of evidence and most are not fully automatic. For example, a recent method utilises coverage across two samples, composition and linkage but through a two-dimensional visualisation approach requiring manual partitioning and validation\cite{albertsen13}. Such an approach is not reproducible and cannot scale to incorporate information from more than two samples. Here we present CONCOCT, a computer program that uses Gaussian mixture models (GMMs)\cite{fraley98} to cluster contigs into genomes based on sequence composition (kmer frequencies) and coverage across multiple samples. To determine the number of clusters we perform model selection using the Bayesian information criteria (BIC)\cite{fraley98}. Since this typically results in slightly more clusters than number of genomes, it is followed by an automatic cluster-merging step that uses information on read-pair linkage within and across clusters.
\section*{Results}

\subsection*{Synthetic Mock Metagenome}

In order to validate CONCOCT we constructed a synthetic mock metagenome data set of 64 samples, each sample comprising random paired-end reads from the same 41 genomes but with different relative frequencies. These frequencies were taken from true abundances of related organisms in 64 human fecal samples determined by 16S rRNA sequencing as part of the Human Microbiome Project\cite{hmp12}. The abundances of the 41 species across the samples are shown in \figref{HMP_Abund}. These reads were then used to generate a co-assembly of all 64 samples giving contigs. A contig is a stretch of consensus sequence generated by overlapping reads. Statistics summarising the co-assembly are given in \tabref{coassembly}. The reads were then mapped back onto the contigs to determine the contig coverages in each of the samples. The coverage is defined as the average number of reads mapping to each position in the contig in that sample. These coverages across samples reflect the abundances of the underlying organisms across the samples and, hence, can be used to disentangle which contig derives from which organism's genome.

The sequence composition of the contigs also contains information, since different organisms have different characteristic signatures of k-mers \cite{sandberg01}, specific n-grams of nucleotide sequence, reflecting mutational biases \cite{knight01}. This forms the basis of most existing non-supervised binning methods \cite{chatterji08,dick09,wang12}. To enable us to incorporate both composition and coverage we developed a unique pre-processing strategy. For each contig we generate a combined profile, joining the coverage and composition vectors together. This is a high dimensional object possessing a total of 65 coverage dimensions - a normalised coverage in each sample and a total coverage - plus 136 k-mer dimensions, if tetramers are used as here. We then perform a principle components analysis (PCA), to reduce the dimensionality of this 201 dimensional combined vector, keeping enough (22) dimensions $D$ to maintain 90\% of the information. The transformed vectors then distill information from both coverage and composition. In \figref{PCASpecies} we plot the 10,458 contigs from the coassembly that exceeded 1000 base pairs (bp) in length in the first two PCA dimensions. The species form distinct clusters even in this two dimensional visualisation, strongly implying that clustering them in the full $D=22$ dimensional space should allow the different species to be easily resolved.

To cluster the contigs we require a method that is unsupervised, operates in any number of dimensions, and can automatically determine the number of clusters present. Model based clustering using an explicit likelihood to describe the data as a mixture of cluster components each described by a separate distribution fulfills these criteria. It requires no human input. The number of dimensions is simply a parameter of the model allowing us to use all $D$ dimensions and model selection based on the BIC can be used to determine cluster number\cite{fraley98}. Then we simply need to determine the form of the distributions describing the clusters, it is apparent from \figref{PCASpecies} that each species forms an ellipsoid in the transformed space, this strongly motivates the use of Gaussian mixture models (GMMs) with full covariance matrices, where each cluster component describes a fully flexible ellipse. This also has the advantage that software packages exist for efficiently fitting these models\cite{scikit-learn}.  

In CONCOCT we couple the above pre-processing strategy with a GMM implementation to automatically cluster metagenome contigs. Applying this to the synthetic community with 64 samples we predicted an optimal cluster number of 56 as the one which minimises the BIC (\figref{MockBIC}). The clusters are visualised in the first two PCA dimensions in \figref{MockClusters}. For the synthetic community we can directly compare to the known assignments of contigs to genomes. We illustrate this graphically as a heat map in \figref{MockConf} and present statistics quantifying the comparison in \tabref{Comp}. We use three statistics precision, recall and the adjusted Rand index (see methods). The precision describes the purity of the clusters, this is remarkably high, 0.97, so that on average a cluster almost entirely contains contigs from the same genome. The recall conversely gives the proportion of each genome that derives from the same cluster, it is how complete the genomes are, here we obtain 0.84, this is less than one because some genomes, as is clear from \figref{MockConf}, are split across multiple clusters. Finally, the adjusted Rand index summarises both precision and recall. For this we obtain 0.83 for the synthetic community.

To illustrate how much additional information the coverage contains over composition alone and to determine how many samples are necessary to resolve a data set of the complexity of the synthetic mock we reran CONCOCT for different number of samples (\figref{MockSS}). The overall accuracy of the clusterings starts to decrease below 16 samples (\figref{MockSS}C), mostly due to a loss of precision although recall is impacted too. Clustering without coverage performs poorly although better than random with an adjusted Rand index of 0.31. 

The basic CONCOCT algorithm performs very well in terms of precision but to improve recall we implemented a further clustering step that utilises read pair linkage information. If a substantial portion of links from contigs in one cluster are to contigs in a different cluster these clusters are merged, provided the two clusters have similar coverage patterns. This reduced the number of clusters to 51 and improving recall to 0.91, without impacting precision, giving a substantially better adjusted Rand index of 0.94 (\tabref{Comp}). 

We also evaluate clusters based on 36 single-copy core genes (SCGs) that are found in almost all known bacterial genomes once (\tabref{SCGs}). The counts of these genes are shown in \figref{MockSCGs} for the clusters without linkage. This suggests that cluster 39 is chimeric which is confirmed by \figref{MockConf}, this cluster contains all the {\it Akkermansia muciniphila} contigs but also 17 contigs from {\it Coriobacterium glomerans}. The other clusters, however, appear pure, some are incomplete though, reflecting the splitting of genomes across clusters mentioned above. Read pair linkage clustering counteracts this problem without negatively impacting purity of the clusters (\figref{MockSCGsL}).

\subsection*{2011 Shiga toxin-producing {\it Escherichia coli} (STEC) O104:H4 outbreak metagenome}

We next applied CONCOCT to a real metagenomic dataset consisting of a total of over 300 million reads from 53 fecal samples taken from individuals suspected of being infected during the 2011 outbreak of Shiga toxin-producing {\it Escherichia coli} (STEC) O104:H4 \cite{loman13}. 43 of these samples derive from individuals that tested positive for STEC by PCR, ten samples were diagnosed as containing other pathogens. Following co-assembly of these reads, 22,585 contigs of length greater than 1,000 bp were obtained. We could taxonomically classify 8,058 of these contigs to species level. A total of 51 species were observed but only 19 with greater than five assigned contigs. This agreed well with the 31 clusters predicted to be optimal by CONCOCT (\figref{STECBIC}). In contrast to other studies of the human gut microbiome, the low diversities observed here are likely because most samples are dominated by \textit{E. coli}. We compared the clusters to the 36\% of contigs which could be classified (\figref{STECConf} and \tabref{Comp}). As for the synthetic mock we obtained a very high precision of 0.94. The recall was somewhat lower, 0.79, probably because in a real community from multiple individuals some species will be split across separate, distantly related strains. In \figref{STECSCGs} we give the frequencies of the SCGs in each cluster wthis suggests that we have 11 complete and fairly pure genomes, with another ten fragments and then some clusters that lack markers. 

An important use of metagenomic data in pathogen discovery is the reconstruction of entire pathogen genomes, without requiring a reference sequence. We expect that the {\it E. coli} outbreak genome will be more abundant in recently infected individuals. Therefore we correlated the time since onset of diarrhea, $ddays$, with the mean log-coverage profiles for the 43 STEC samples.  These are obtained by back-transforming the means of each cluster component (\tabref{Corr}). There are ten clusters with a false discovery rate of less than 10\%. The relative abundance of these clusters across samples ordered by $ddays$ is shown in the top panel of \figref{STECMapCluster}. The two most significant clusters, $k=7$ and $k=18$, are both strongly negatively correlated with $ddays$, they are also predominately made up of contigs assigned to {\it Escherichia coli}. They are therefore candidates for the outbreak genome. We can verify this because the outbreak genome is known\cite{loman13}, in \tabref{Map} we give the number of contigs from each cluster that map either to parts of the {\it Escherichia coli} O104:H4 genome that are specific to that strain or core {\it E. coli} genes. 94.8\% of the mapped contigs derive from clusters 7 and 18. These two clusters together represent the outbreak genome. Many more contigs mapping to the {\it E. coli} core genome derive from cluster 18 and a higher proportion of outbreak-specific contigs derive from cluster 7 (\tabref{Map} and \figref{STECMapGenome}). The separation of the outbreak strain into core and outbreak portions probably reflects both the greater variation in nucleotide composition in non-core bacterial genes and differences in coverage where other non-outbreak {\it E. coli} strains are present. From the lower panel of \figref{STECMapCluster} it is apparent that there is a greater variability in cluster 7 than 18. This leads to the single genome being best described by two components with different variances. 

We have focused on the outbreak {\it E. coli} clusters but the clustering approach in CONCOCT facilitates a whole community analysis, this is evident from \figref{STECMapCluster}, there are two further clusters negatively correlated with $ddays$, clusters 8 and 1, very few contigs from the former can be classified. Cluster 1 is from {\it Enterococcus faecium}, potentially evidence of this organism increasing in abundance in association with STEC infection. The other clusters are all positively associated with $ddays$. These are organisms that may be important in the recovery from STEC infection, they include, cluster 29 - {\it Acidaminococcus intestini}, and cluster 16 - {\it Ruminococcus sp. SR1/5}. The other clusters are unknown Bacteroides or Clostridia species. Note that as the clusters move from negative to positive correlation with $ddays$ their means move to the right in the transformed lower panel of \figref{STECMapCluster}. The most important PCA dimension, PCA1, is separating outbreak associated organisms from the healthy commensal microbiota. Finally, applying linkage clustering reduces the cluster number from 31 to 29 and improves recall to 0.87, giving an adjusted Rand index of 0.82 (see \tabref{Comp} and \figref{STECSCGsL} for SCGs). In fact, two of the clusters joined were the {\it Escherichia coli} clusters 7 and 18 recovering the near-complete outbreak genome as a single cluster. 

The CONCOCT software is downloadable from \verb=https://github.com/BinPro/CONCOCT= and test data from \verb=https://github.com/BinPro/CONCOCT-test-data=.

\section*{Discussion}

\section*{Materials and Methods}


\subsection*{Synthetic HMP data set}

We based the simulation of our mock data set on 64 16S rRNA samples from the HMP project \cite{hmp12}. These samples were denoised with the AmpliconNoise pipeline\cite{quince11} and OTUs constructed at 3\% to approximate species. This generated a total 6,839 OTUs with known relative abundance profiles across the samples. After filtering out OTUs with a total of less than 50 counts summed across samples, we used BLAST to match the remaining OTU sequences against the NCBI whole genome database\cite{altschul90}. We were able to identify distinct organisms for 41 OTUs (with a blast identity of $>$ 95\%). If multiple OTUs were assigned to the same organism, then we chose the OTU with higher abundance across all samples. We then renormalised the abundance profiles for these 41 genomes (see \figref{HMP_Abund}) and identified the genomes present in the individual communities. The 41 genomes derived from 39 different species and 21 genera. Each sample comprised between 7 and 27 non-zero OTUs. For each of the 41 organisms in our mock database we compiled all chromosomes and plasmids available on NCBI. Unknown nucleotides (N) were deleted from the sequences for the purpose of the simulation.

We based the number of simulated reads on the yield of approximately 1/2 flowcell (4 lanes) of an Illumina HiSeq 2500 high output run. We assumed that each lane yields 188 million read pairs, so for each of our 64 samples we simulated 11.75 million paired-end read pairs. Reads were generated by sampling randomly across the genomes present in a sample according to their relative abundance and then sampling position uniformly at random within the selected genome. Our read simulation programme utilises position and nucleotide specific substitution, insertion and deletion patterns. We inferred these error profiles based on a real data set, where a diverse \textit{in vivo} mock community was prepared with the TruSeq library preparations method and sequenced on a HiSeq. We also inferred the fragment size distribution from this data set and used it for the simulation of the paired-end 2x100bp reads. The read simulation programme outputs reads in fasta format which we converted into a pseudo fastq format for the downstream analysis assuming uniform quality scores. The read names contain information on the genome from which the respective read originated for the subsequent validation of the CONCOCT algorithm.

\subsection*{{\it E. coli} O104:H4 outbreak}

In addition to the synthetic community we considered a real data set. This consisted of over 300 million reads filtered for human DNA from 53 metagenomic datasets from individuals presenting during the 2011 outbreak of Shiga-toxin producing {\it Escherichia coli} (STEC) O104:H4 \cite{loman13}. 43 of these samples were from STEC infected individuals, with the remaining ones from patients with clinical diagnoses of {\it Clostridium difficile}, {\it Salmonella enterica} and {\it Campylobacter jejuni}.

\subsection*{Coassembly and mapping}

For both the simulated and real data a coassembly of reads from all samples was performed. Ray version 2.1.0\cite{boisvert12} was used to generate the coassemblies, because it is able to handle large datasets by distributing the computation over multiple nodes and is specifically designed to handle metagenomics data. For the {\it E. coli} O104:H4 outbreak only 10\% of the reads were used in the coassembly in order to complete on a machine with 256GB RAM with a k-mer length of 31. For the synthetic data, the k-mer length was set to 41, this took six hours using 1,024 cores on a Cray XE6 system. 

To determine coverage of the coassembly per sample, the reads were mapped back with bwa version 0.7.5a-r405\cite{li2010fast} using the mem algorithm. The coverages were subsequently computed with BEDTools\cite{quinlan10}. The linkage between contigs was determined using the \verb=bam_to_linkage.py= script, which is part of the CONCOCT package. The script searches for paired reads in a bam file that align to the tips of two different contigs, where a tip was chosen to be 500 bases. Since the synthetic coassembly community is based on simulated reads, it was furthermore possible to determine the origin of each read mapping to a certain contig. Each contig can then be assigned to a genome based on a majority vote. This script can also be found in CONCOCT: \verb=contig_read_count_per_genome.py=. The {\it E. coli} contigs were taxonomically classified by searching against the NCBI database using \verb=TAXAassign v0.4=\cite{taxaassign}. This programs uses BLAST\cite{altschul90} to search for matches from the NCBI nucleotide database that are within a given identity and query coverage. Taxonomy is then assigned based on the top $n$ hits that match these criteria using a consensus approach i.e. we assign at a given level if at least 80\% of the hits at that level have the same taxa. We used 90\% for both identity and coverage, and the top 100 sequences. These values were chosen to ensure a stringency sufficient that we could be confident in species level assignments.

\subsection*{Preprocessing}

Prior to preprocessing contigs are filtered by length, only contigs greater than a minimum size are used. This is an adjustable parameter but was fixed at 1000 base pairs throughout this study. Each filtered contig indexed $i = 1,\ldots,N$ is represented by a coverage vector and a composition vector. The coverage vector is simply the average number of reads per base from each of $M$ samples, indexed $j = 1,\ldots,M$, mapping to that contig. We will denote the coverage vector for each contig by $\mathbf{Y_i} = \left(Y_{i,1},\ldots,Y_{i,M}\right)$. The composition vector contains the frequency for each k-mer and its reverse complement in that contig. In all the results presented here a k-mer frequency of 4 was used although CONCOCT can accept any k-mer length as a parameter. The frequencies are combined with complements since sequencing is bidirectional. The dimension $V$ of the composition is 136 for tetramers due to palindromic k-mers. We denote the composition vector for each contig by $\mathbf{Z_i} = \left(Z_{i,1},\ldots,Z_{i,V}\right)$. Prior to normalisation we added a small pseudo-count to both coverage and composition vectors. This removes non-zero entries, necessary to allow the log-transform below. It is equivalent to assuming a uniform Dirichlet prior on the relative frequencies. For the composition we simply add a single count to each k-mer $Z'_{i,j} = Z_{i,j} + 1$ but for the coverage we imagine mapping an extra read of length 100bp to each contig in each sample $Y'_{i,j} = Y_{i,j} + 100/L_i$ where $L_i$ is the contig length.

Coverage vectors are normalised, firstly over samples, so that different read numbers from a sample are accounted for:
\begin{equation}
Y''_{i,j} = \frac{Y'_{i,j}}{\sum_{k = 1}^{N} Y'_{k,j}},
\end{equation}
and over contigs to give coverage profiles $\mathbf{p}$. This normalises for coverage variation within a genome ensuring that this does not mask co-occurrence of contigs:
\begin{equation}
p_{i,j} = \frac{Y''_{i,j}}{\sum_{k=1}^{M}Y''_{i,k}}.
\end{equation} 
However, the total coverage does contain further information that may potentially discriminate organisms, so we keep this as an additional variable $Y''_{i,.}=\sum_{k=1}^{M}Y''_{i,k}$.
We also normalise composition to give composition profiles $\mathbf{q}$. This accounts for different contig lengths:
\begin{equation}
q_{i,j} = \frac{Z'_{i,j}}{\sum_{k=1}^{V}Z'_{i,k}}.
\end{equation}  
These two vectors and the total coverage are joined together and log-transformed to give a combined log-profile $\mathbf{x_i} = \left[\log(p_{i,1}),\ldots,\log(p_{i,M}),\log(Y''_{i,.}),\log(q_{i,1}),\ldots,\log(q_{i,V}) \right]$ of dimension $E = M + V + 1$. This vector then represents both coverage and composition, the transform expands the domain of the normalised variables to the negative half-space, and gives improved results over not-transforming or alternatives such as the square root. Finally, a dimensionality reduction using principle components analysis, implemented as a singular value decomposition, was performed on the $N \times E$ matrix of log-profiles $\mathcal{X}$ with rows corresponding to the vectors $\mathbf{x_i}$ and thus elements, $x_{i,j}$. The number of components, $D$, necessary to explain 90\% of the variance in the data were kept. This reduces the dimensionality of the clustering problem whilst keeping the majority of the coverage and composition information. We will denote the transformed data $N \times D$ matrix by $\mathcal{V}$ with row vectors $\mathbf{v_i}$ for each contig $i$.
 
\subsection*{Gaussian mixture model (GMM)}

To cluster contigs into bins we use a Gaussian mixture model. This defines the data likelihood as a sum of $K$ Gaussians:
\begin{equation}
\mathcal{L}\left(\mathcal{V}|K,\pi_k,\mathbf{\mu_k},\Sigma^2_k\right) = \sum_{i = 1}^{N} \sum_{k = 1}^{K} \pi_k \mathcal{N}\left(\mathbf{v}_i | \mathbf{\mu_k},\Sigma^2_k \right),
\end{equation}
where $\mathcal{N}$ denotes a Gaussian or normal distribution. Each of these Gaussian components corresponds to a different cluster. They are each parameterised by their own mean vectors, $\mathbf{\mu_k}$ and standard deviations, $\Sigma^2_k$ for which full covariance structures were used. This allows a very flexible specification of the cluster as an ellipse in the $D$ dimensional reduced space. The proportions of each mixture component in the data set are given by the $K$ dimensional vector $\mathbf{\pi} = (\pi_1,\ldots,\pi_K)$. These are effectively prior probabilities that a contig will derive from a cluster before considering its data vector. We fit the GMM using the scikit-learn Python library \cite{scikit-learn}. This uses an expectation-maximisation (EM) algorithm to determine both the mixture model parameters and, crucially, the responsibilities, $z_{i,k}$, which can be viewed as the probability that contig $i$ derives from cluster, $k$. For most of the statistics, we use a hard assignment, obtained by assigning contig $i$ to the cluster for which the $z_{i,k}$ is maximised denoting this $\gamma_i \in 1,\ldots,K$.

To determine the correct number of clusters, $K$ in a sample we adopt a model selection approach based on the Bayesian information criterion (BIC). The BIC is defined as: 
\begin{equation}
BIC = -2ln\left[\mathcal{L}\left(\mathcal{V}|K,\pi_k,\mathbf{\mu_k},\Sigma^2_k\right)\right] + p.ln(N)
\end{equation} 
where $p$ is the total number of parameters, $p = K.D(D + 3)/2$. The BIC therefore penalises badly fitting models and parameter number simultaneously. We try a range of $K$ values and select the model that minimises the BIC. This is a standard approach to model selection for GMMs \cite{fraley98}. It is also useful to consider the cluster means in the original space, transforming the $\mathbf{\mu_k}$ vectors from the $D$ dimensional reduced space back to the original $E$ dimensions of the combined log-profiles, we will denote these back-transformed vectors as $\mathbf{\mu_k'}$. In particular, the first $M + 1$ dimensions of these vectors give the mean cluster profile of the component which can be correlated with sample-specific meta-data.

\subsection*{Evaluating clusterings by comparison to known genome assignments}

For the synthetic communities we know the genome each contig derived from, defining this as the genome from which the majority of reads mapping to that contig derived. We can view these as class labels. These class assignments represent the idealised grouping, a perfect clustering would predict a $K$ equal to the number of classes or genomes, 41, and with each cluster purely composed of contigs from one of the genomes. For the {\it E. coli} O104:H4 outbreak data we do not know the true species assignments for the majority of contigs but we do for those that we could unambiguously assign with the \verb=TAXAAssign= script, 8,058 out of the 22,585 contigs with length $>$ 1000bp, we can use these labelled contig for evaluation there too. In reality, we will never obtain a perfect clustering, we therefore need a statistic to determine how far from that perfect grouping a clustering is. For this the Rand index is an intuitive solution, it considers pairs of elements, if a pair of elements deriving from the same class are placed in the same cluster then this is considered a true positive, denote the number of such pairs as $TP$. Similarly if a pair of elements deriving from different classes are placed in different clusters then this is considered a true negative, $TN$. The Rand index lying between 0 and 1, is simply the number of corrects pairs, $TN + TP$, divided by the total number of pairs $N \choose 2$. However, given a random classification and a random clustering we would expect a non-zero Rand index just by chance. The adjusted Rand index accounts for this by subtracting the expected value given fixed class and cluster sizes\cite{hubert85} and normalising so that values are still smaller than equal to 1 which indicates a perfect clustering. In addition to the adjusted Rand index we used two further measures that help to understand in what way a clustering is deviating from the classification. The first is the recall, here we calculate the number of contigs from each species or class that are placed in the same cluster, sum over all classes and divide by $N$. 
This indicates how complete the genome bins are. The second is the precision, which inversely is calculated by summing the contigs in each cluster that derive from the same species and dividing by $N$, it indicates how pure the clusters are. These statistics are generated by the script \verb=Validate.pl= distributed in CONCOCT from the cluster and class assignments.

\subsection*{Evaluating clusters with single-copy core genes (SCGs)}

Evaluating clusters with single-copy core genes is an alternative way of evaluating cluster completeness and purity. We utilize housekeeping genes that typically occur in single copies in microbial genomes.  To select appropriate genes we downloaded all complete microbial genomes from NCBI (August 25, 2013) and selected one random genome for each genus. We counted the frequency of each Cluster of Orthologous Groups (COGs)\cite{tatusov97} of genes in these 525 genomes.  Since there are very few COGs that occur once in every genome we instead applied the more relaxed criteria of being present in greater than 97.0\% of the genomes and having an average frequency of less than 1.03. This resulted in 36 COGs. Twenty-seven of these are shared with the list of 40 COGs that was selected in a similar way in an earlier study\cite{ciccarelli06}. The script \verb=Validate_scg.pl= generates a table of counts for these COGs (or another list of genes provided by the user) within the different clusters output by CONCOCT. The 36 selected COGs are given in \tabref{SCGs}.

\subsection*{Incorporating linkage}

The clusters produce by the GMM are of a high precision mostly containing contigs from just a single species but some species are split between multiple clusters. To address this we incorporate a further source of information, paired-end linkage. The standard Illumina sequencing protocol can generate reads from either end of a longer fragment, these reads are known as paired. In some cases the paired reads will not map to the same contig but different, this is referred to as linkage. This suggests that these two contigs in fact derived from the same genome sequence. We use this information to devise a further clustering algorithm, but one which is a hierarchical agglomerative clusterer, joining the clusters generated by the GMM if there are sufficient links between but with the restriction that the mean coverage profiles of the clusters are similar enough. We begin by defining a symmetric $N \times N$ linkage matrix, $\mathcal{L},$ with elements $l_{i,j}$ corresponding to the number of linked reads between two contigs $i$ and $j$. If the cluster of contigs $k$ with members $U_k$ comprising all contigs with $\gamma_i = k$ derives from the same genome as another cluster $l$ with members $U_l$ then we would expect a sufficiently large portion of the links formed by $U_k$ to join with $U_l$. We quantify this by defining a $K \times K$ link transition matrix $\mathcal{T}$ with elements $t_{k,l}$:
\begin{equation}
t_{k,l} = \frac{\sum_{i \in U_{k}}\sum_{j \in U_l}H\left[l_{i,j} - l_m\right]}{\sum_{i \in U_{k}} \sum_{j=1}^{N} H\left[l_{i,j} - l_m\right]}
\end{equation}
where $H\left[l_{i,j} - l_m\right]$ is a Heaviside function such that:
\begin{equation}
  H\left[l_{i,j} - l_m\right]=\begin{cases}
    0, & \text{if $l_{i,j} - l_m < 0$}.\\
    1, & \text{otherwise}.
  \end{cases}
\end{equation}
The constant $l_m$ acts as a threshold above which we consider two contigs to have enough linked reads to count as `linked'. This helps eliminate noise due to badly mapped reads or chimeric contigs, for all results here we used $l_m = 10$.
All linked pairs are given equal weighting through the Heaviside function to account for different coverage levels. The $\mathcal{T}$ matrix gives the proportion of links from cluster $k$ to cluster $l$ normalised by the total number of links associated with $k$. We then calculate the mean coverage profile of each cluster:
\begin{equation} 
\langle \mathbf{p_k} \rangle = \frac{\sum_{i \in U_k} p_{i,j}}{|U_k|}, 
\end{equation}
and a $K \times K$ coverage overlap matrix $\mathcal{O}$ which gives the degree of overlap between any two cluster coverage profiles:
\begin{equation}
o_{k,l} = \sum_{j=1}^M \textrm{min}(\langle p_{k,j} \rangle, \langle p_{l,j}\rangle).
\end{equation}
The clustering algorithm itself consists of the following steps:
\begin{enumerate}
\item Join the two clusters $k',l'$ with largest off-diagonal value $t_{k,l}$ but only if $o_{k,l} > o_m$

\item Update the $\mathcal{T}$ and $\mathcal{O}$ matrices by creating a new aggregate cluster comprising the union of $U_{k'}$ and $U_{l'}$
Repeat until largest $t_{k,l} < t_m$. 
\end{enumerate}
Clusters are joined provided they are sufficiently correlated until the majority of links are within rather than between clusters. We used values of $t_m = 0.05$ and $o_m = 0.8$.

\section*{Acknowledgments}

This research arose out of a workshop funded through the COST project ES1103 graciously hosted by Pedro Fernandes at the Instituto Gulbenkian de Ci\^{e}ncia in Lisbon. 
This work was funded by the Swedish Research Councils VR [grant 2011-5689], FORMAS [grant 2009-1174] and EU-FP7 [grant BLUEPRINT] through grants to A.F.A. C.Q. is funded by an EPSRC Career Acceleration Fellowship - EP/H003851/1. M.S. is supported by Unilever R\&D Port Sunlight, Bebington, United Kingdom. The computations were performed on resources provided by the Swedish National Infrastructure for Computing (SNIC) at PDC Centre for High Performance Computing (PDC-HPC) and Uppsala Multidisciplinary Center for Advanced Computational Science (UPPMAX).

\bibliography{CONCOCT}
 
\clearpage
\section*{Figures}

\begin{figure}[p]
\begin{center}
\includegraphics[width=4in]{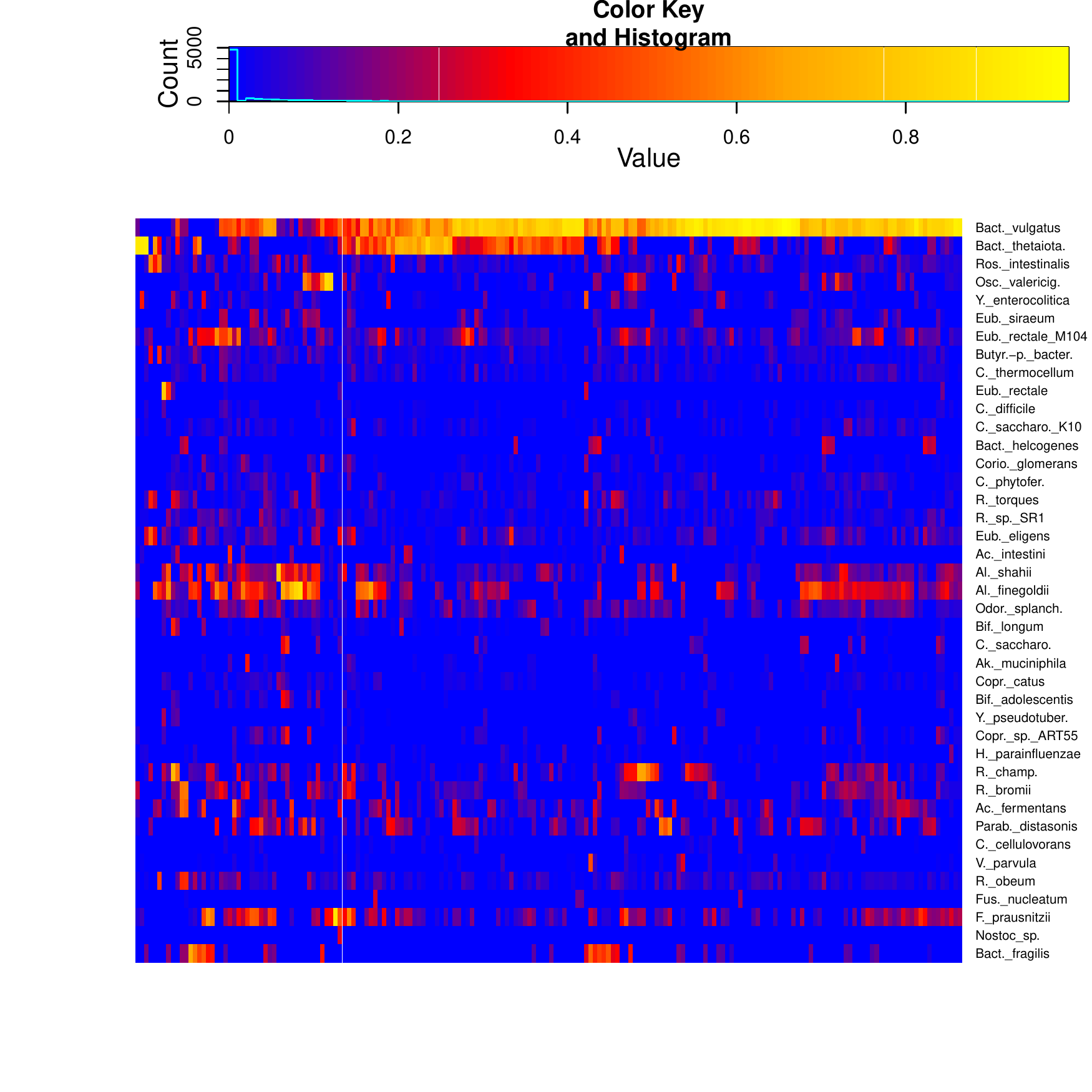}
\end{center}
\caption{
{\bf Heat map of the relative abundances of the 41 genomes in the synthetic mock community distributed across the 64 HMP samples.} The species and samples have been positioned according to similarity. The relative abundances have been square root transformed to emphasise rare species and the inset scale should be interpreted accordingly.
}
\figlabel{HMP_Abund}
\end{figure}

\begin{figure}[p]
\begin{center}
\includegraphics[width=4in]{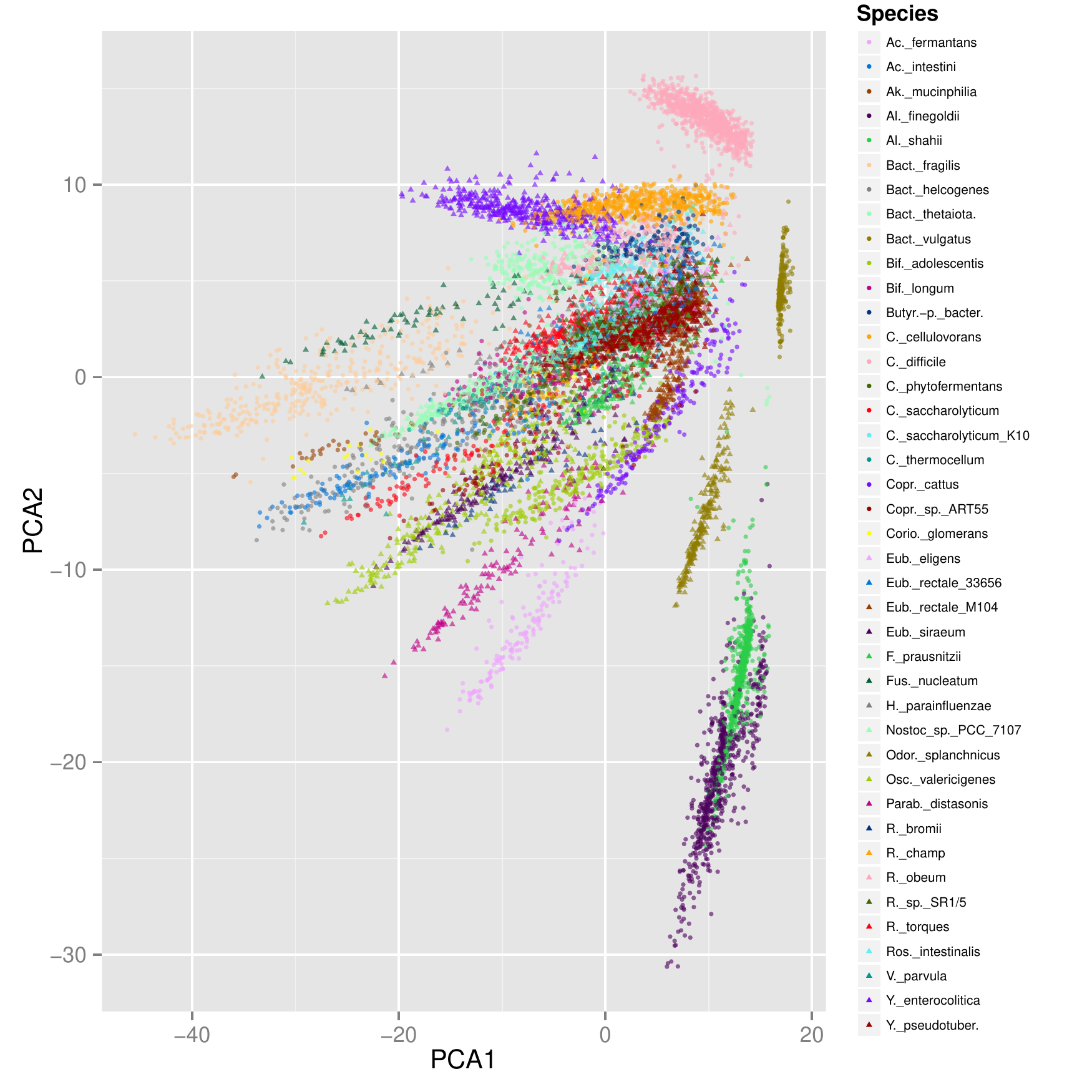}
\end{center}
\caption{
{\bf PCA plot of the mock metagenome species.} The 10,458 synthetic community contigs of length $>$ 1000bp plotted in the first two PCA dimensions with the 41 different species discriminated.}
\figlabel{PCASpecies}
\end{figure}

\begin{figure}[p]
\begin{center}
\includegraphics[width=4in]{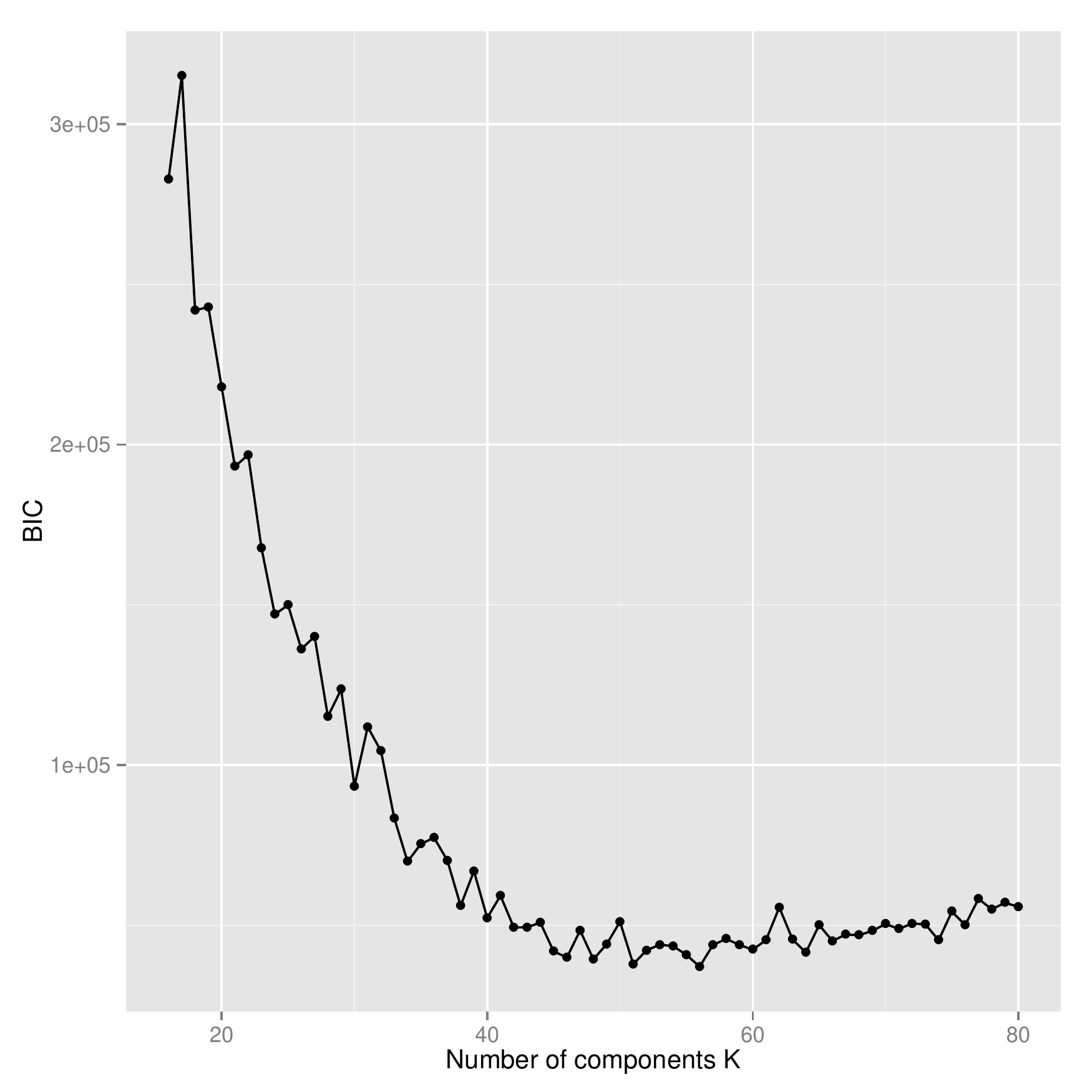}
\end{center}
\caption{
{\bf The Bayesian information criterion (BIC) as a function of cluster number $K$ for the synthetic mock community.} Results are the minimum from five runs with independent random starting conditions. The optimal $K$ with the smallest BIC is 56.}
\figlabel{MockBIC}
\end{figure}

\begin{figure}[p]
\begin{center}
\includegraphics[width=4in]{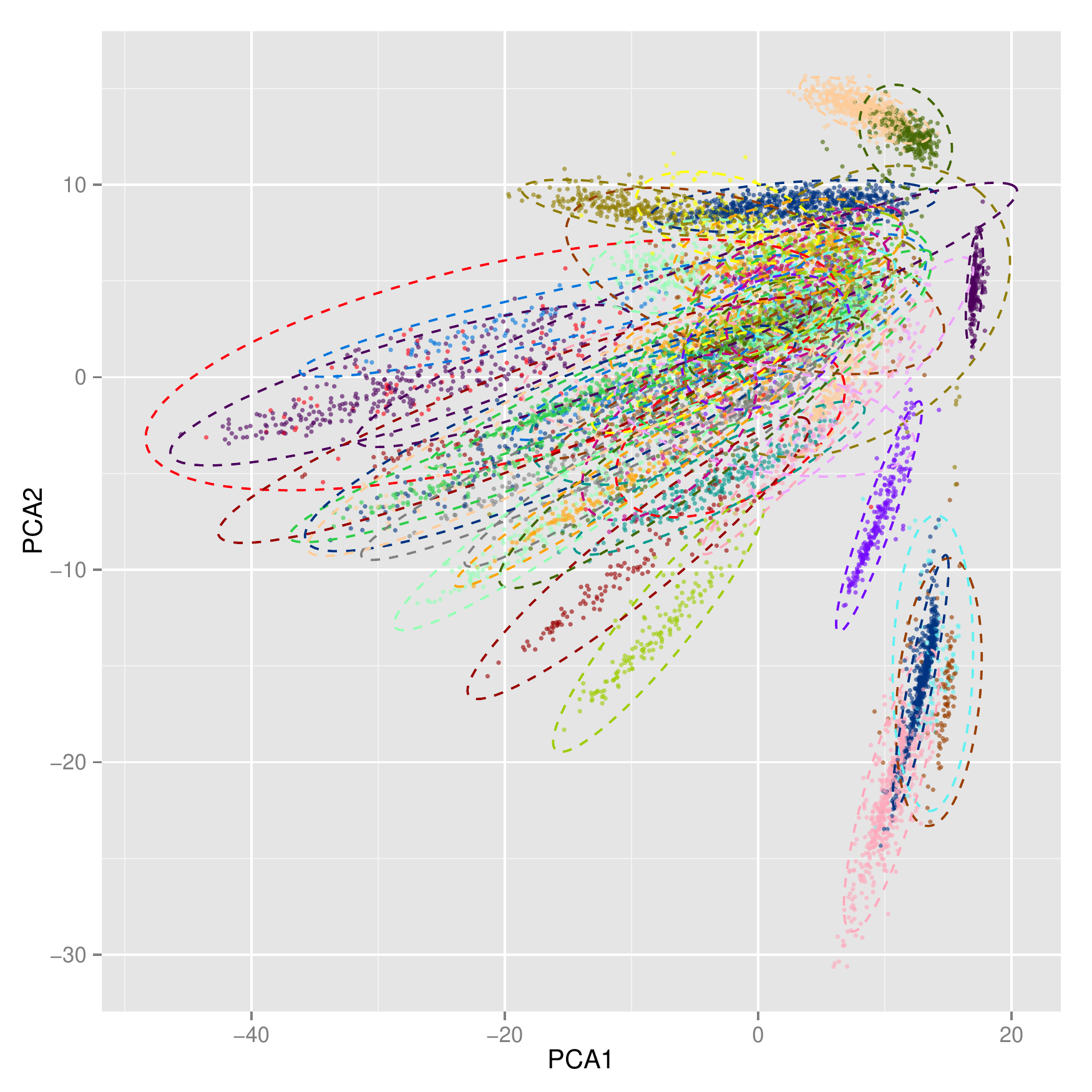}
\end{center}
\caption{
{\bf PCA plot of the mock metagenome clusters.} The 10,458 synthetic community contigs of length $>$ 1000bp plotted in the first two PCA dimensions with the 56 different contig clusters discriminated.}
\figlabel{MockClusters}
\end{figure}

\begin{figure}[p]
\begin{center}
\includegraphics[width=4in]{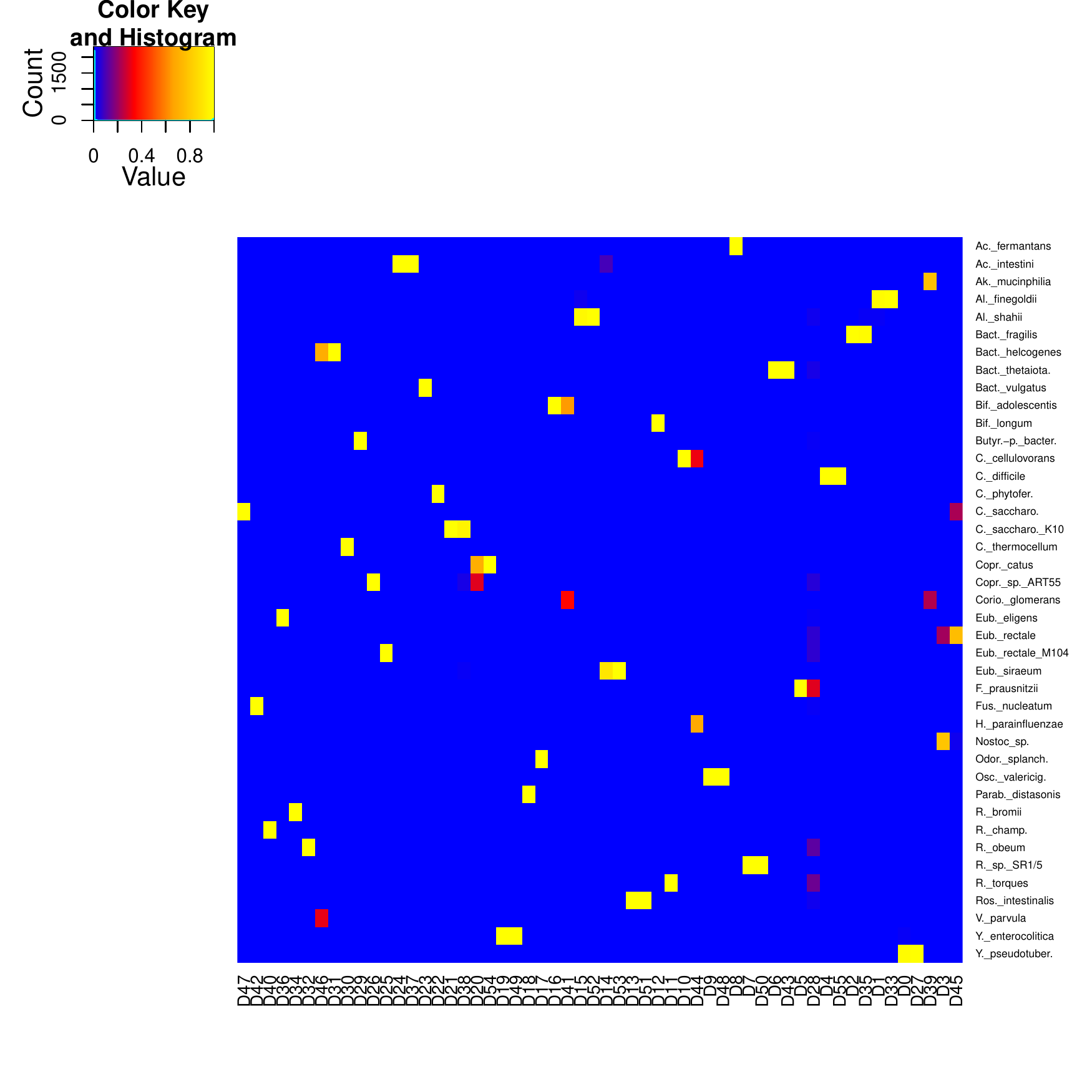}
\end{center}
\caption{
{\bf Confusion matrix for the mock metagenome.} A heatmap visualisation of the confusion matrix comparing the CONCOCT contig clusterings (without linkage) for the optimal 56 cluster solution with the species assignments for the synthetic mock contigs. Each column is a cluster named D$k$ where $k$ is the cluster index. The rows correspond to the species and the intensities reflect the proportion of each cluster deriving from each species.}
\figlabel{MockConf}
\end{figure}

\begin{figure}[p]
\begin{center}
\includegraphics[width=4in]{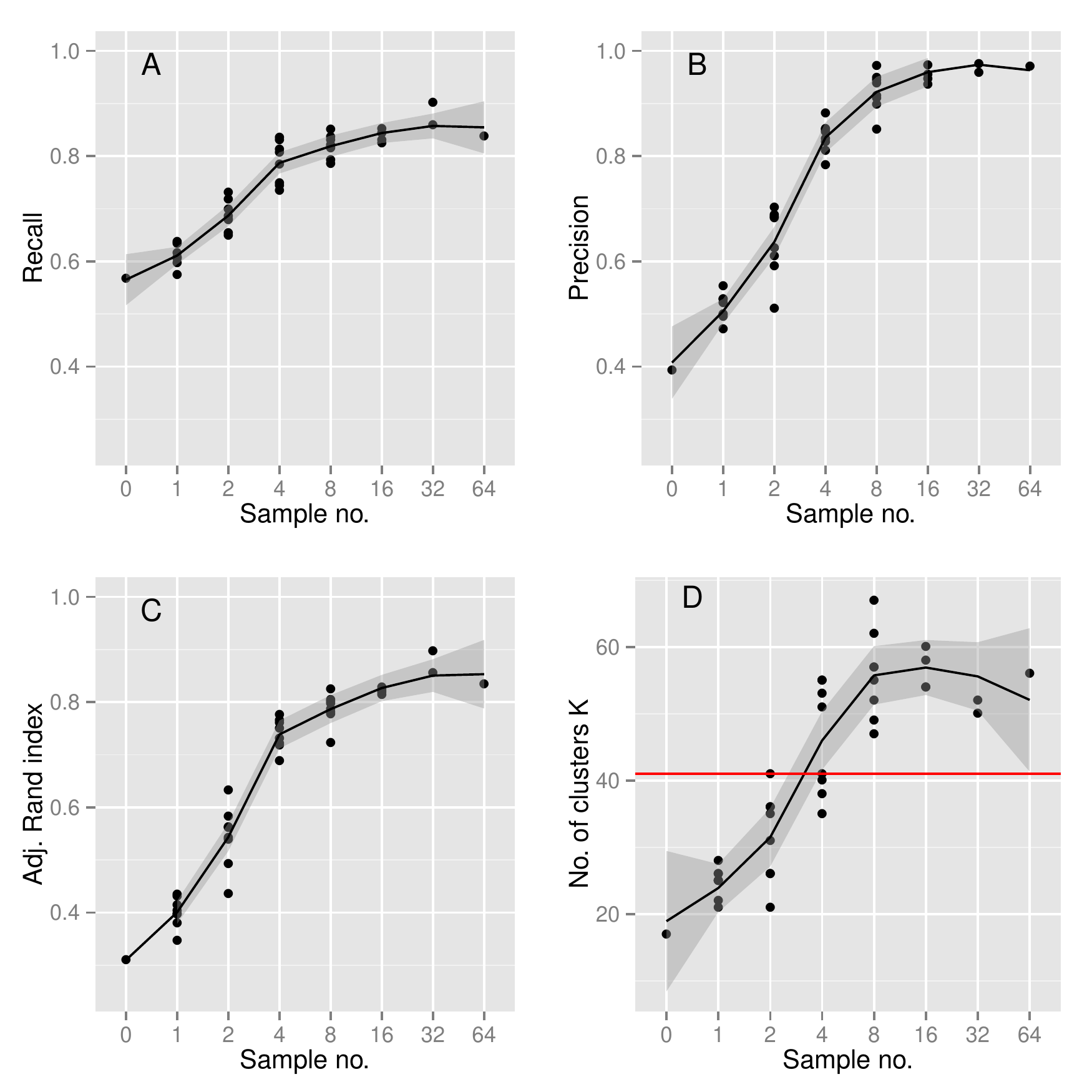}
\end{center}
\caption{
{\bf Impact of sample number on accuracy of synthetic mock community clustering.} CONCOCT was run with variable sample numbers the optimal K determined (D) and recall (A), precision (B) and adjusted Rand index (C) calculated.}
\figlabel{MockSS}
\end{figure}

\begin{figure}[p]
\begin{center}
\includegraphics[width=4in]{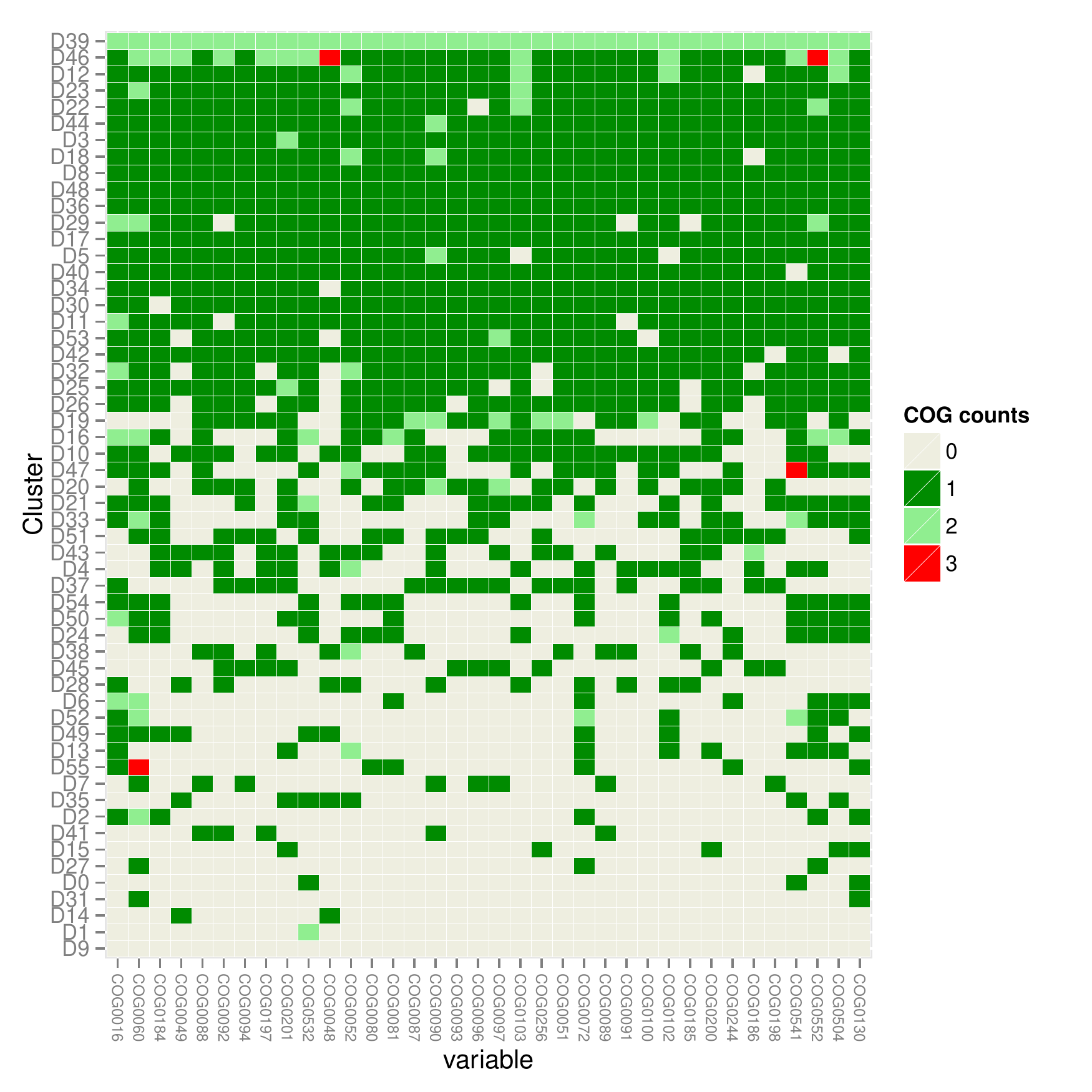}
\end{center}
\caption{
{\bf Single-copy core gene frequencies in the mock metagenome clusters.} A heatmap visualisation of the number of single-copy core genes in each cluster for the optimal 56 cluster solution generated by CONCOCT without linkage applied to the synthetic mock community.}
\figlabel{MockSCGs}
\end{figure}

\begin{figure}[p]
\begin{center}
\includegraphics[width=4in]{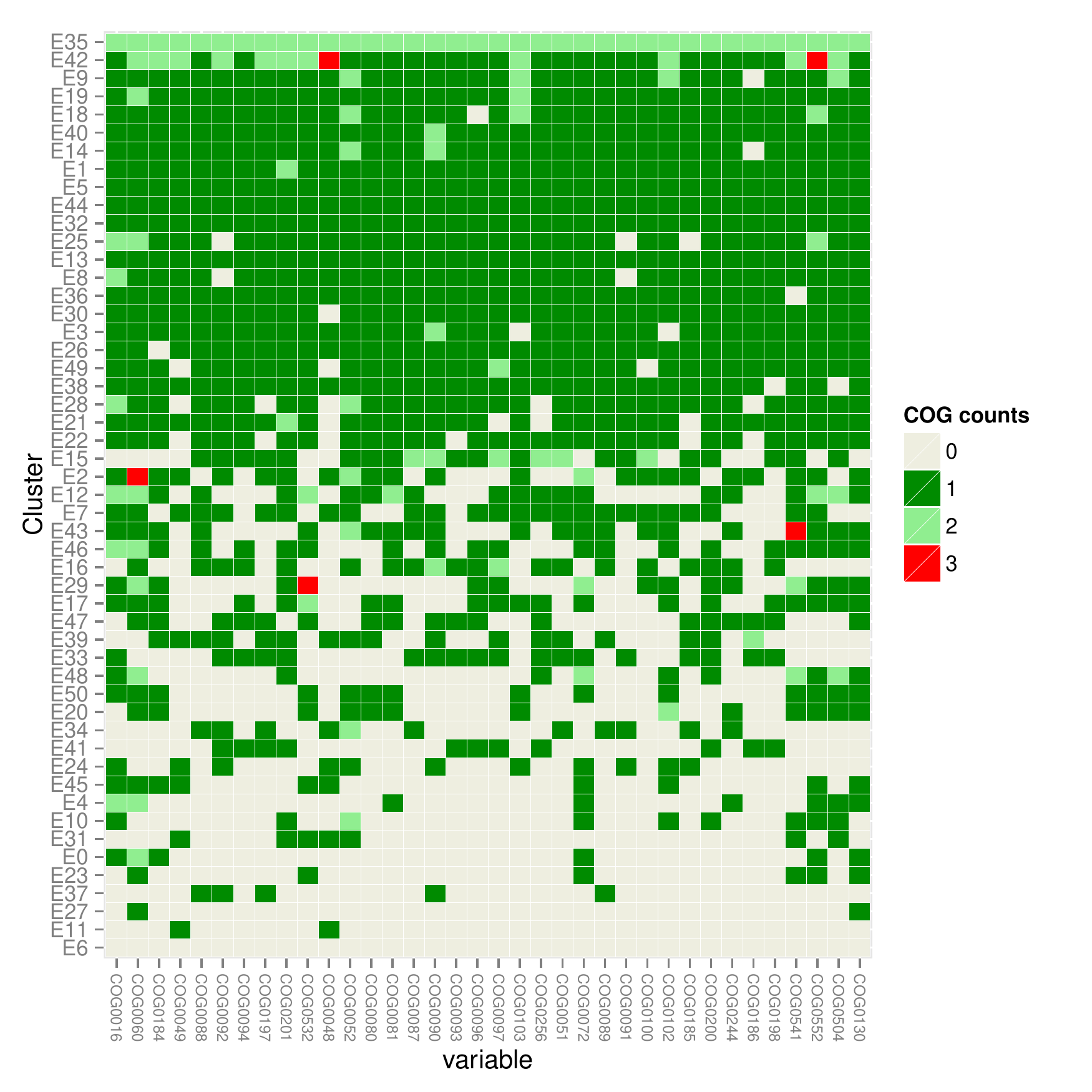}
\end{center}
\caption{
{\bf Single-copy core gene frequencies in the mock metagenome clusters.} A heatmap visualisation of the number of single-copy core genes in each cluster for the optimal 51 cluster solution generated by CONCOCT with linkage applied to the synthetic mock community.}
\figlabel{MockSCGsL}
\end{figure}

\begin{figure}[p]
\begin{center}
\includegraphics[width=4in]{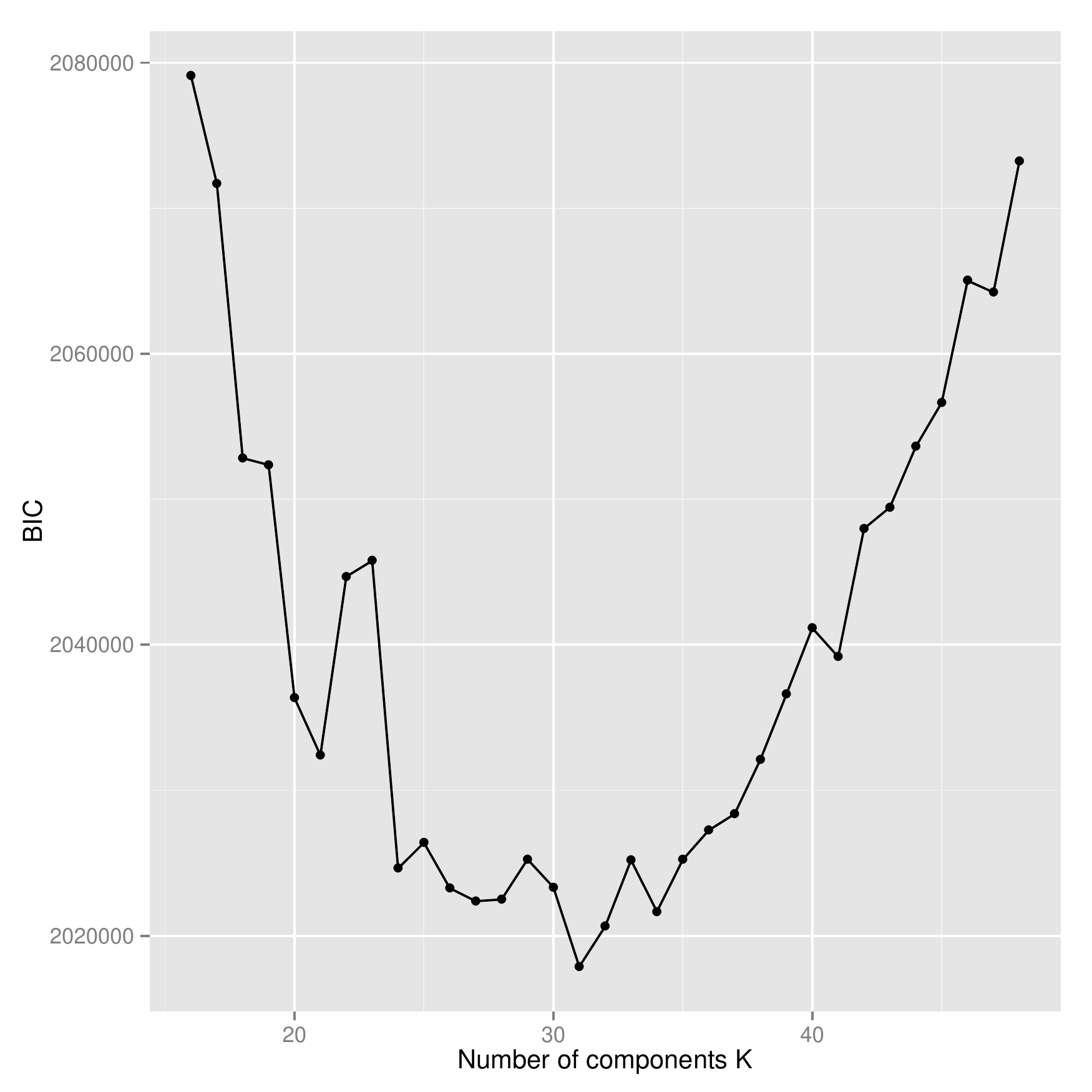}
\end{center}
\caption{
{\bf BIC as a function of $K$ for the {\it Escherichia coli} (STEC) O104:H4 data set.} The Bayesian information criterion (BIC) as a function of cluster number $K$ for the {\it Escherichia coli} (STEC) O104:H4 data set. Results are the minimum from five runs with independent random starting conditions. The optimal $K$ with the smallest BIC is 31.}
\figlabel{STECBIC}
\end{figure}

\begin{figure}[p]
\begin{center}
\includegraphics[width=4in]{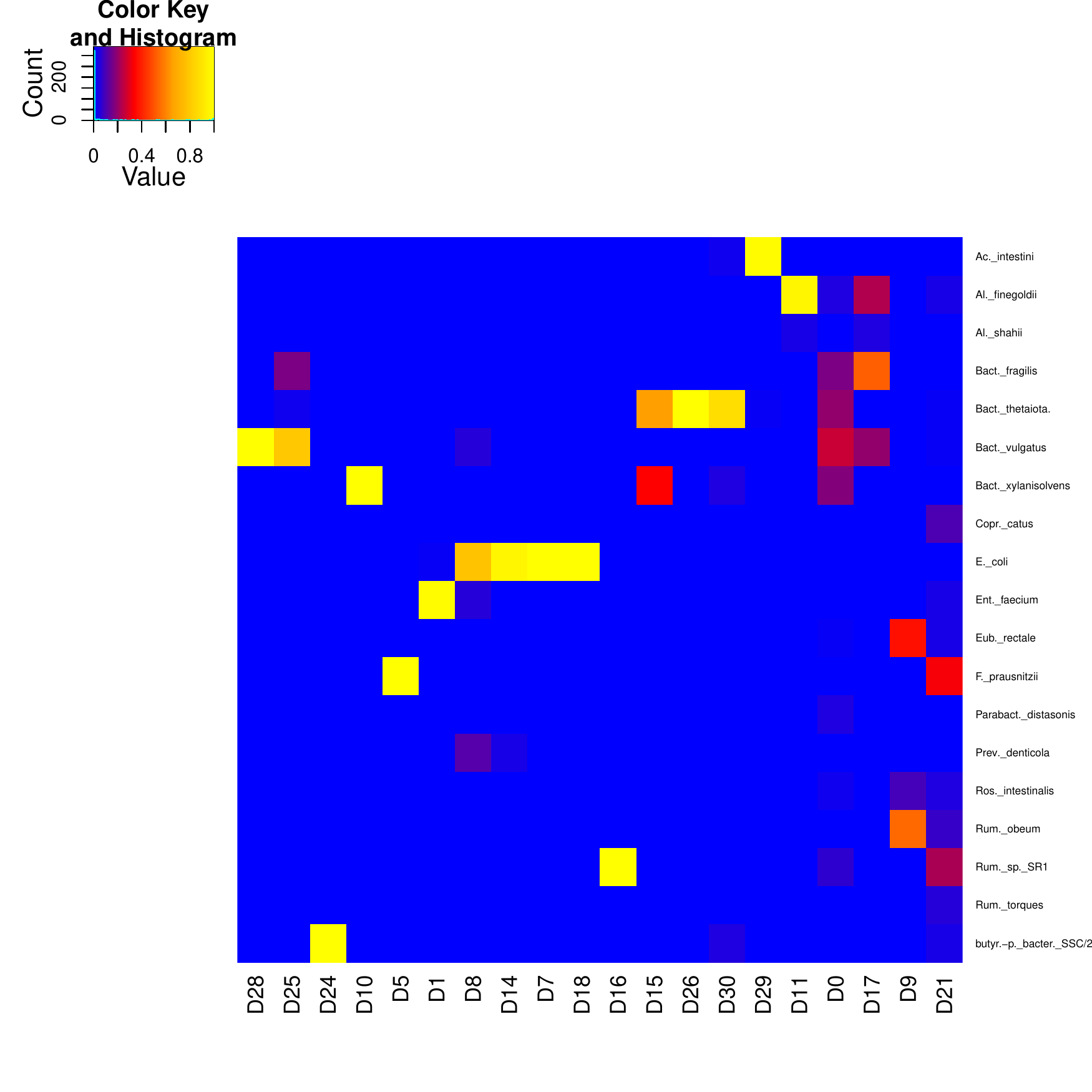}
\end{center}
\caption{{\bf Confusion matrix for the {\it Escherichia coli} (STEC) O104:H4 metagenome.} A heatmap visualisation of the confusion matrix comparing the CONCOCT contig clusterings (without linkage) for the optimal 31 cluster solution of the shiga-toxigenic {\it Escherichia coli} (STEC) O104:H4 outbreak. Each column is a cluster named D$k$ where $k$ is the cluster index. The rows correspond to the species assignments, the intensities reflect the proportion of each cluster deriving from each species. Only clusters and taxa with greater than five labelled representatives are shown.}
\figlabel{STECConf}
\end{figure}

\begin{figure}[p]
\begin{center}
\includegraphics[width=4in]{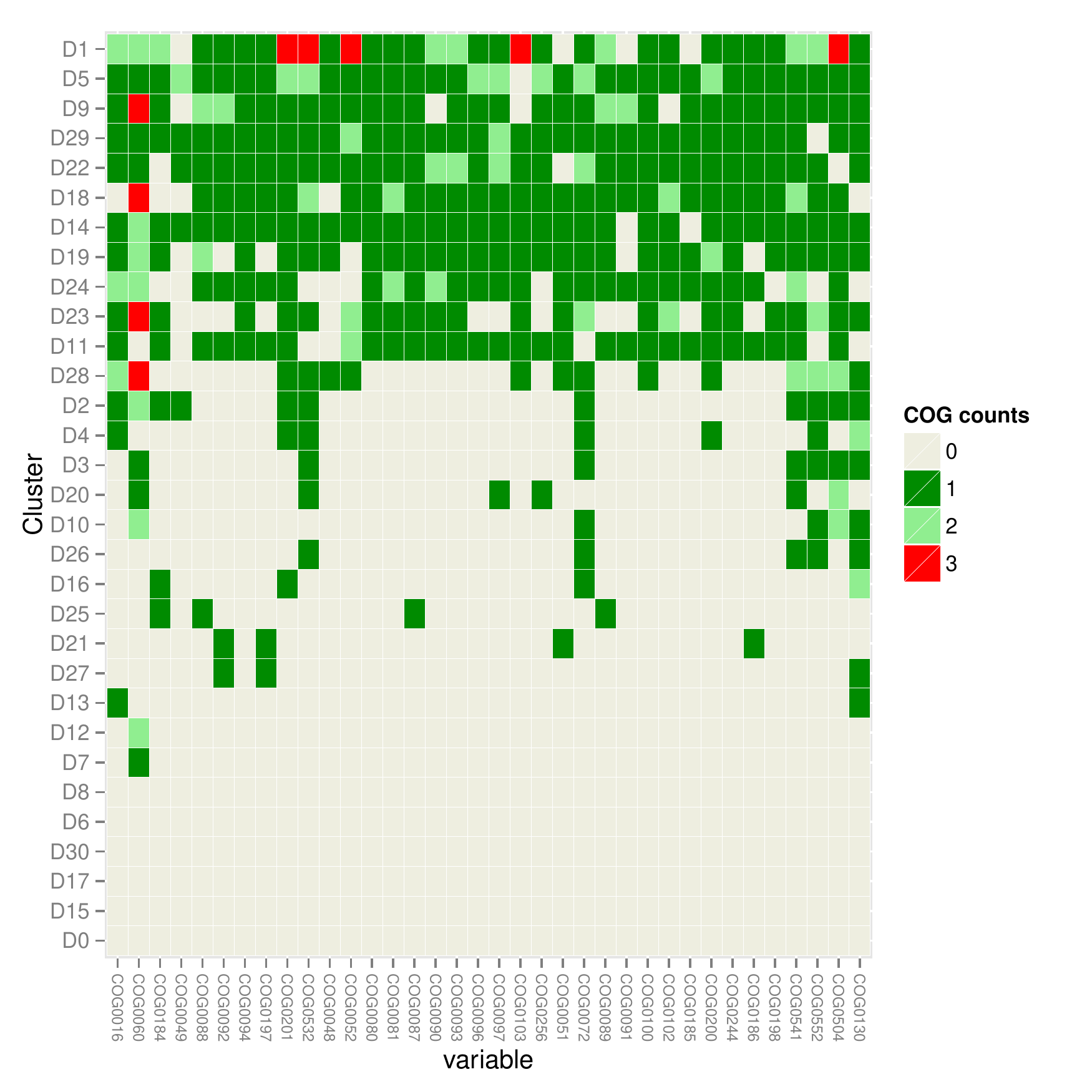}
\end{center}
\caption{{\bf Single-copy core gene frequencies in the {\it Escherichia coli} (STEC) O104:H4 metagenome clusters.} A heatmap visualisation of the number of single-copy core genes in each cluster for the optimal 31 cluster solution without linkage of CONCOCT applied to the shiga-toxigenic {\it Escherichia coli} (STEC) O104:H4 outbreak.}
\figlabel{STECSCGs}
\end{figure}

\begin{figure}[p]
\begin{center}
\includegraphics[width=4in]{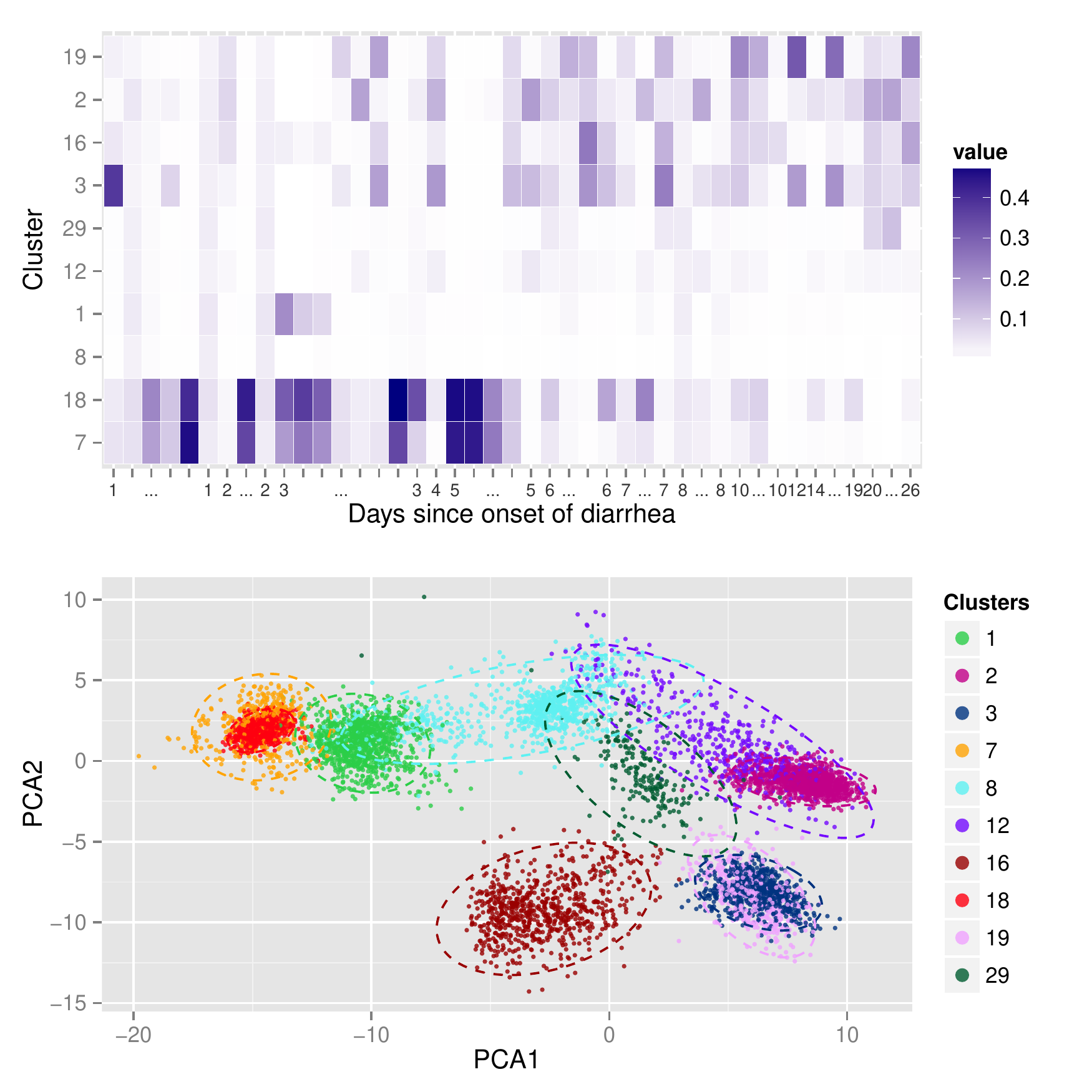}
\end{center}
\caption{{\bf Visualisation of key cluster abundances and location in PCA map from the {\it Escherichia coli} (STEC) O104:H4 metagenome.} Top panel: heat map giving the relative abundance of the ten clusters that correlated significantly with days since onset of diarrhea, $ddays$, in each of the 43 STEC positive samples. The STEC samples have been ordered by $ddays$. The clusters are ordered bottom to top from negative to positive correlation coefficients (\tabref{Corr}). Bottom panel: the same ten clusters plotted in the first two PCA components of the transformed combined log coverage composition profiles.}
\figlabel{STECMapCluster}
\end{figure}

\begin{figure}[p]
\begin{center}
\includegraphics[width=4in]{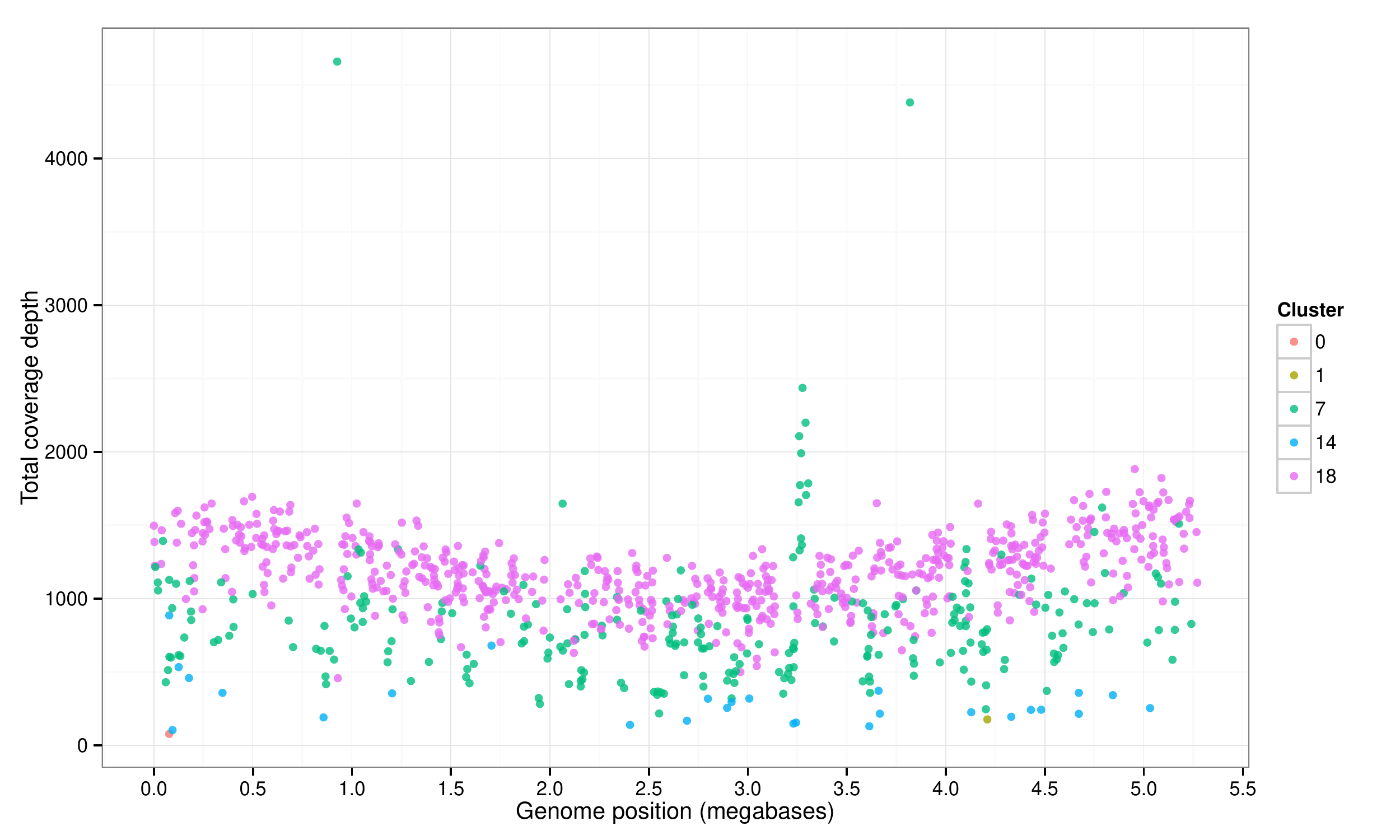}
\end{center}
\caption{{\bf Mapping of contigs to known outbreak genome.} The mapping of contigs to the known {\it Escherichia coli} (STEC) O104:H4 outbreak genome with cluster discriminated by colour and total coverage across all samples shown on the y-axis.}
\figlabel{STECMapGenome}
\end{figure} 

\begin{figure}[p]
\begin{center}
\includegraphics[width=4in]{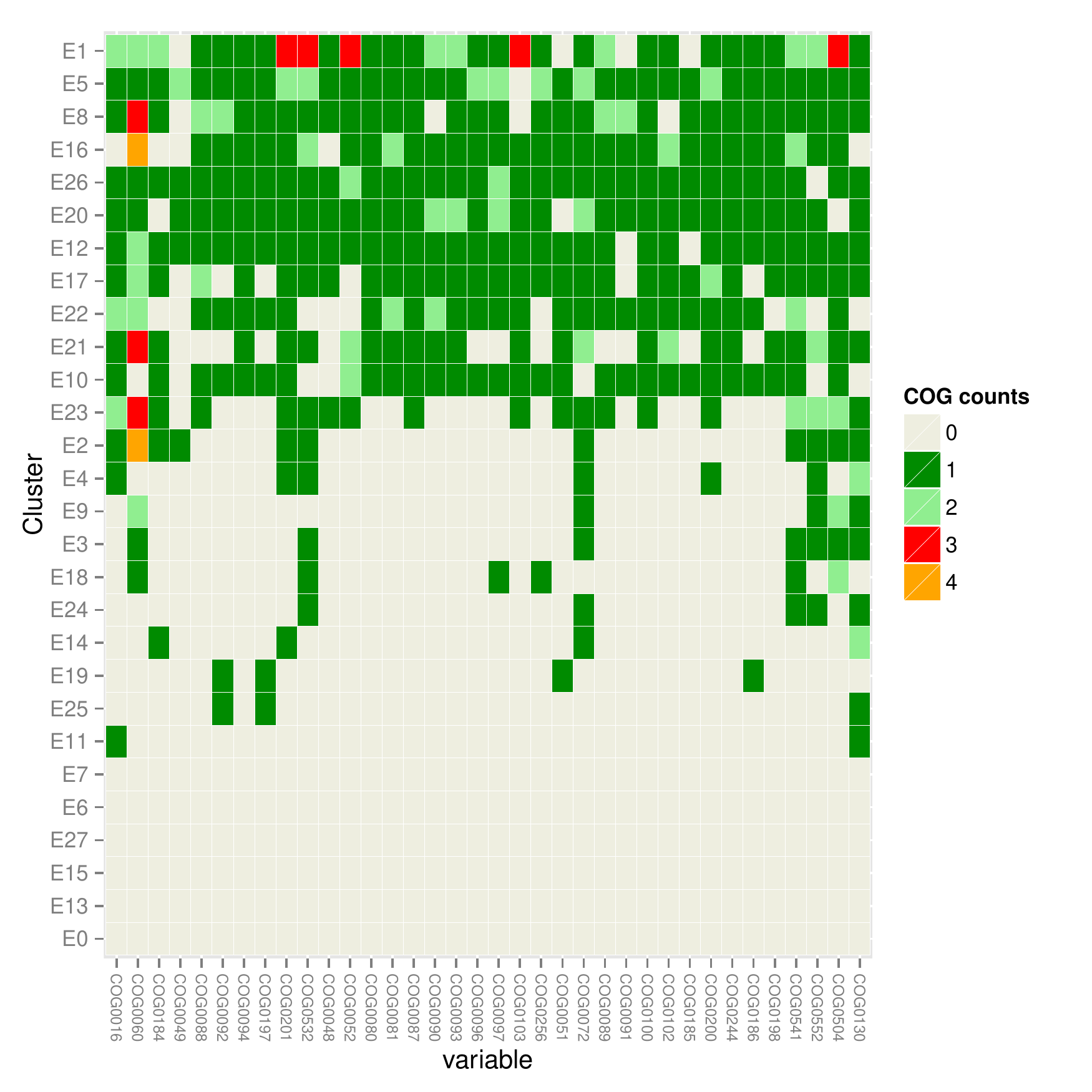}
\end{center}
\caption{{\bf Single-copy core gene frequencies in the {\it Escherichia coli} (STEC) O104:H4 metagenome clusters following application of linkage.} A heatmap visualisation of the number of single-copy core genes in each cluster for the optimal 28 cluster solution with linkage of CONCOCT applied to the shiga-toxigenic {\it Escherichia coli} (STEC) O104:H4 outbreak.}
\figlabel{STECSCGsL}
\end{figure}

\clearpage

\section*{Tables}

\begin{table}[p]
\begin{tabular}{|l|c|c|c|c|c|c|}
\hline
Sample & k-mer & no. reads & no. contigs& no. contigs $>$ 1000bp& $N50_{length}$& \%age ref.\\
\hline
Mock & 41 & 1.504$\times10^9$ & 34,986 & 10,459 & 32,551 & 88.07 \\
\hline
E. coli & 31 &33,882,279 & 154,360       &  22,585      &  1,587  & ---\\
\hline
\end{tabular}
\caption {Co-assembly statistics for the synthetic mock and the {\it Escherichia coli O104:H4} outbreak.  Ray version 2.1.0\cite{boisvert12} was used to generate the coassemblies. The E. coli assembly was a random subsample of 10\% of the reads in the study\cite{loman13}.}
\tablabel{coassembly}
\end {table}

\begin{table}[p]
\begin{tabular}{|l|c|c|c|c|c|c|c|}
\hline
Sample & $N$ & $N'$& $S$ & $K$ & Rec. & Prec. & Adj. Rand \\
\hline
Mock & 10,459 & 10,459 & 41 & 56 & 0.84	& 0.97	& 0.83\\
+ linkage & 10,459 & 10,459 & 41 & 51 & 0.91 & 0.97 & 0.94\\
\hline
E. coli & 22,585 & 8,058 & 51 & 31 & 0.79 & 0.94 & 0.73\\
+ linkage & 22,585 & 8,058 & 51 & 28 & 0.87 & 0.94 & 0.82\\
\hline
\end{tabular}
\caption {Cluster validation statistics for the synthetic mock community and the {\it Escherichia coli O104:H4} outbreak. The number of contigs with length $> 1000bp$ clustered is $N$, the number with class labels $N'$, the number of distinct classes i.e. species $S$, the optimal cluster number $K$, Rec. is the recall, Prec. the precision and Adj. Rand the adjusted Rand index.} 
\tablabel{Comp}
\end {table}

\clearpage

\begin{table}[p]
\tiny
\begin{tabular}{|l|c|c|c|}
\hline
COG	& Name & Presence (\%)	& Mean frequency \\
\hline
COG0016	& Phenylalanyl-tRNA synthetase alpha subunit & 99.6	& 1.02 \\
COG0060	& Isoleucyl-tRNA synthetase & 99.6	& 1.01\\
COG0184	& Ribosomal protein S15P/S13E & 99.6	& 1\\
COG0049	& Ribosomal protein S7 & 99.4	& 1\\
COG0088	& Ribosomal protein L4 & 99.4	& 1\\
COG0092	& Ribosomal protein S3 & 99.4 & 1.01\\
COG0094	& Ribosomal protein L5 RPL11 & 99.4 & 1\\
COG0197	& Ribosomal protein L16/L10E & 99.4 & 	1\\
COG0201	& Preprotein translocase subunit SecY & 99.4 & 	1.01\\
COG0532	& Translation initiation factor 2 & 99.4 & 	1.01\\
COG0048	& Ribosomal protein S12 & 99.2  & 	1\\
COG0052	& Ribosomal protein S2 & 99.2 & 	1\\
COG0080	& Ribosomal protein L11 & 99.2 & 	1\\
COG0081	& Ribosomal protein L1 & 99.2 & 	1\\
COG0087	& Ribosomal protein L3 & 99.2 & 	1\\
COG0090	& Ribosomal protein L2 & 99.2 & 	1\\
COG0093	& Ribosomal protein L14 & 99.2 & 	1\\
COG0096	& Ribosomal protein S8 & 99.2 & 	1\\
COG0097	& Ribosomal protein L6P/L9E & 99.2 & 	1\\
COG0103	& Ribosomal protein S9 & 99.2 & 	1\\
COG0256	& Ribosomal protein L18 & 99.2 & 	1\\
COG0051	& Ribosomal protein S10 & 99 & 	1.02\\
COG0072	& Phenylalanyl-tRNA synthetase beta subunit & 99 & 	1\\
COG0089	& Ribosomal protein L23 & 99 & 	1\\
COG0091	& Ribosomal protein L22 & 99 & 	1\\
COG0100	& Ribosomal protein S11 & 99 & 	1\\
COG0102	& Ribosomal protein L13 & 99 & 	1\\
COG0185	& Ribosomal protein S19 & 99 & 	1\\
COG0200	& Ribosomal protein L15 & 99 & 	0.99\\
COG0244	& Ribosomal protein L10 & 99 & 	0.99\\
COG0186	& Ribosomal protein S17 & 98.9 & 	1\\
COG0198	& Ribosomal protein L24 & 98.5 & 	0.99\\
COG0541	& Signal recognition particle GTPase & 98.5 & 	1\\
COG0552	& Signal recognition particle GTPase  & 98.5 & 	1\\
COG0504	& CTP synthase (UTP-ammonia lyase) & 97.9 & 	1.02\\
COG0130	& Pseudouridine synthase & 97.5 & 	0.99\\
\hline
\end{tabular}
\caption {The list of single-copy core genes (SCGs) used for evaluating cluster completeness and purity, the percentage of genomes in which they are present, and their mean frequencies within genomes. Calculations were based on 525 microbial genomes, each representing a unique genus.} 
\tablabel{SCGs}
\end {table}

\begin{table}[p]
\begin{tabular}{|l|c|c|c|}
\hline
Cluster - D$k$ & $p_a$ & $p$ & $r$ \\
\hline
D7 & 0.001394846 & 4.499503e-05 & -0.580448290\\
D18 & 0.014409298 & 9.296321e-04 & -0.486898053\\
D19 & 0.022560790 & 2.585918e-03  & 0.448066807\\
D2 & 0.022560790 & 2.911070e-03  & 0.443265420 \\
D8 & 0.090729544 & 2.026110e-02 & -0.352944330 \\
D1 & 0.090729544 & 2.239690e-02 & -0.347547700\\
D16 & 0.090729544 & 2.655809e-02  & 0.338165455\\
D3 & 0.090729544 & 2.733903e-02  & 0.336543155\\
D29 & 0.090729544 & 2.855075e-02 & 0.334100554\\
D12 & 0.090729544 & 2.926759e-02 & 0.332695738\\
\hline
\end{tabular}
\caption {Correlation between days since onset of diarrhea $ddays_i$ for each sample $i$ and the first $M$ components of the back-transformed cluster coverage profiles $\mathbf{\mu_k'}$. The $p$ and $r$ values give the significance and correlation respectively, between the mean log coverage profile at a sample $\mu'_{k,i}$ for a cluster denoted D$k$ and $ddays_i$ using Pearson's test. The $p_a$ denote Benjamini-Hochberg corrected false discovery rates. Only clusters with $p_a < 0.1$ are shown. Only the 43 STEC positive samples were used in this analysis.} 
\tablabel{Corr}
\end {table}

\begin{table}[p]
\begin{tabular}{|l|c|c|c|c|c|c|}
\hline
Cluster - D$k$ & D0 & D1 & D7 & D8 & D14 & D18\\
\hline
No. contigs $|U_k|$ & 843 & 958 & 437 &862 & 732 & 953\\ 
Outbreak specific & 1 & 1 & 205 & 0 & 10 & 18\\
Core & 0 & 0 & 203 & 3 & 65 & 935\\
Plamid pAA & 0 & 0 & 22 & 0 & 0 & 0\\
Plasmid pESBL & 0 & 0 & 16 & 0 & 0 & 0\\
\hline
\end{tabular}
\caption {Comparison of clusters to the {\it E coli} core and outbreak {\it E. coli} O104:H4 STEC 280 genomes. Each of the 22,585 contigs of greater than 1000bp were mapped to the two genomes and outbreak plasmids using MUMMER. Only six clusters had any contigs mapping and for these the numbers classified to either core {\it E. coli} genes present in both genomes or outbreak specific genes found only in {\it E. coli} O104:H4 are shown. In addition, to the number mapping to two outbreak associated clusters. The total number of contigs in each cluster $|U_k|$ is also given.} 
\tablabel{Map}
\end {table}

\end{document}